\documentclass[aps,prc,reprint,showpacs,floatfix,nofootinbib]{revtex4-1}
\usepackage{CJK}
\usepackage{graphicx,amsmath}
\usepackage{aas_macros}
\usepackage{multirow}

\begin{document}
\begin{CJK*}{GB}{} 
\title{Critical assessment of nuclear sensitivity metrics for the $r$-process}
\author{Zachary Shand}
\affiliation{Department of Physics \& Astronomy, University of Calgary, Calgary, AB, T2N 1N4, Canada}
\author{Rachid Ouyed}
\affiliation{Department of Physics \& Astronomy, University of Calgary, Calgary, AB, T2N 1N4, Canada}
\author{Nico Koning}
\affiliation{Department of Physics \& Astronomy, University of Calgary, Calgary, AB, T2N 1N4, Canada}

\author{Iris Dillmann}
\affiliation{Department of Physics and Astronomy, University of Victoria, Victoria, BC, V8P 5C2, Canada}
\affiliation{TRIUMF, 4004 Wesbrook Mall, Vancouver, BC V6T 2A3, Canada}

\author{Reiner Kr\"{u}cken}
\affiliation{Department of Physics and Astronomy, University of British Columbia, Vancouver, BC, V6T 1Z4}
\affiliation{TRIUMF, 4004 Wesbrook Mall, Vancouver, BC V6T 2A3, Canada}


\author{Prashanth Jaikumar}
\affiliation{Department of Physics \& Astronomy, California State University Long Beach, 1250 Belfower Blvd., Long Beach CA 90840 USA}

\begin{abstract}
Any simulation of the $r$-process is affected by uncertainties in our present knowledge of nuclear physics quantities and astrophysical conditions. It is common to quantify the impact of these uncertainties through a global sensitivity metric, which is then used to identify specific nuclides that would be most worthwhile to measure experimentally. Using descriptive statistics, we assess a set of metrics used in previous sensitivity studies, as well as a new logarithmic measure. For certain neutron-rich nuclides lying near the r-process path for the typical hot-wind scenario, we find opposing conclusions on their relative sensitivity implied by different metrics, although they all generally agree which ones are the most sensitive nuclei. The underlying reason is that sensitivity metrics which simply sum over variations in the $r$-process distribution depend on the scaling used in the baseline, which often varies between simulations. We show that normalization of the abundances causes changes in 
the reported sensitivity factors and recommend reporting a minimized 
$F$ statistic in addition to a scale estimation for rough calibration 
to be used when comparing tables of sensitivity factors from different studies. 
\end{abstract}

\maketitle

\end{CJK*}

\section{Introduction}
About half of all the stable nuclei heavier than iron are produced by the mechanism of rapid neutron capture, or the $r$-process~\cite{1957RvMP...29..547B, 1957AJ.....62....9C}, which occurs in explosive neutron-rich astrophysical environments. Obtaining a better fit to the solar system's isotopic abundances of heavy elements (within astrophysical uncertainties) gives confidence that we have identified the primary site of $r$-process nucleosynthesis. Parameterized studies of neutron-rich flows and a growing set of observations from metal-poor stars suggest that no proposed site can produce the entire $r$-process from the first peak around $A$=80 to the third peak at $A$=195~\cite{2003ApJ...588.1099Q, doi:10.1146/annurev.astro.46.060407.145207}.  For a complete understanding of the origin of heavy elements, theoretical simulations of the $r$-process are essential in discriminating between the several proposed astrophysical sites, but are faced with modeling uncertainties in the nuclear physics inputs for neutron-rich nuclei. Rare-isotope measurements at existing and upcoming experimental facilities can help reduce these uncertainties in $r$-process simulations. This requires that key isotopes for the $r$-process in experimentally accessible regions of the nuclide chart be identified for measurements. To accomplish this, quantitative metrics, called ``sensitivity factors,'' have been developed~\citep{PhysRevC.92.035807,2012EPJA...48..184B,2012PhRvC..86c5803M}. 

Nuclear mass models are crucial to the $r$-process, since they affect neutron separation energies ($S_n$), $\beta$-decay Q-values ($Q_{\beta}$), 
$\beta$-decay half-lives (T$_\textrm{1/2}$), neutron-capture cross 
sections ($\sigma$) etc., which are all 
important nuclear input parameters to a full network calculation. Even in the classical waiting-point approximation, the use of different nuclear mass models with variations in the predicted shell structure near the magic numbers alters the neutron separation energy (i.e. the $r$-process path) and the final calculated abundance~\cite{2015PhRvC..92c5807M}. 
Although we will not achieve a complete measured range up to the drip line, as the number of experimentally measured masses increases we can hope that the 
predictive power of the nuclear mass models will improve~\cite{2014PhRvC..90a7302S}. 

To quantify how the uncertainty in nuclear properties propagates to the $r$-process abundance pattern, a global sensitivity measure
(or ``impact parameter") $F$ is utilized in some studies, which is examined in detail in this paper (see \citep{PhysRevC.92.035807,2012EPJA...48..184B,2012PhRvC..86c5803M} for various definitions of $F$). Based on this metric, key pieces of data are identified near the neutron closed shells and the precursors of the rare-earth peak which are most influential in generating the overall abundance pattern~\cite{2014EPJWC..6607024S}. While this metric, and its variations in the series of papers on sensitivity studies (see Ref.~\cite{Mumpower201686} for a review), is a simple way to capture the impact of variations in the nuclear parameters (locally or globally), distilling the sensitivity to a single number is fraught with erroneous conclusions on the relative importance of some isotopes, as we demonstrate in this paper. The freedom to scale or normalize the simulated r-process pattern in an arbitrary way leads to opposing conclusions on the sensitivity depending on the metric used. 

The purpose of this work is two-fold: (i) to show that application of the currently existing metrics lead to varying conclusions on the relative importance of certain nuclides in producing the best fit to the $r$-process abundance pattern; (ii) introduce a statistical significance to the $F$-metric that takes the arbitrariness in baseline normalization into account, so that conclusions on the sensitivity are more refined and specific to the particular metric used. 

This is the first step towards a more universal definition of sensitivity for $r$-process studies. The paper is organized as follows: In Sec.~\ref{sec:simulation}, we set up our $r$-process waiting-point simulation with the SiRop code and calculate the variations in the abundance using different measures due to changes in the nuclear mass model. In Sec.~\ref{sec:metric-compare}, we compare the performance of four different metrics (two absolute, one relative and one log-difference) and highlight differing conclusions on the sensitivity of the $r$-process pattern to nuclear masses around the $A$=130 peak. In Sec.~\ref{sec:normalization-calibration}, we 
describe the effect of scaling and normalization, followed by our conclusions in Sec.~\ref{sec:conclude}.

\section{Simulation Parameters}
\label{sec:simulation}
We generate our $r$-process simulation data using an extension of the $r$-Java 2.0 code~\cite{2014A&A...568A..97K,2014arXiv1402.3824K} that now includes a graphical-user interface (GUI) module for sensitivity runs. This first-of-its-kind code, ``SiRop", allows users to run the $r$-process simulation and output sensitivity metrics in a single Java-based application in order to represent and analyze changes in the abundance curves.
The data used to compare the sensitivity studies was computed in the waiting point approximation for fast 
computation of data to test and compare each metric. The results of the baseline and varied simulations is 
shown in Fig.~\ref{fig:simulationAbundances}. For the baseline, we used a parametric trajectory for the density of $\rho = \frac{\rho_0}{(1+t/2\tau)^2}$ 
where $\rho_0$ = $10^{11} \textrm{~g\,cm}^{-3}$ and $\tau$=$0.001$ sec, and with an initial temperature  $T$=$3\times 10^9 \textrm{K}$. The trajectory ensures validity of the waiting point approximation and production of $r$-process elements including the third peak at $A$=195. 

The initial isotopic 
composition of the environment was set to 50\% $^{70}$Fe by mass (i.e. $X_{70\textrm{Fe}}$=$0.5$) corresponding to an initial 
neutron-to-seed ratio of 62. The code was run until the neutron-to-seed ratio 
dropped below one at which point the temperature and neutron number density of the 
simulation were $2.7\times10^9$~K and $4.64\times10^{28} \textrm{~cm}^{-3}$.  These values 
were used to calculate the waiting point population coefficients displayed in 
Fig.~\ref{fig:metricCharts} as circles. The varied simulations consisted of changes in masses of a single isotope in a grid 
spanning from $^{128}$Cd to $^{139}$Te. These isotopes were chosen as they are in the predicted r-process path and match the 
isotopes selected in \citet{2009PhRvC..79d5809S}. For each isotope, the simulation was run twice corresponding to a decrease and increase in 
the isotopes mass by $0.0005\%$ which corresponds to a change of approximately $\pm$0.6~MeV, where this value corresponds to the average deviation of 
mass models from experimental values. 

\section{Metric performance comparison}
\label{sec:metric-compare}
\begin{figure*}
\centering
\includegraphics[width=\linewidth]{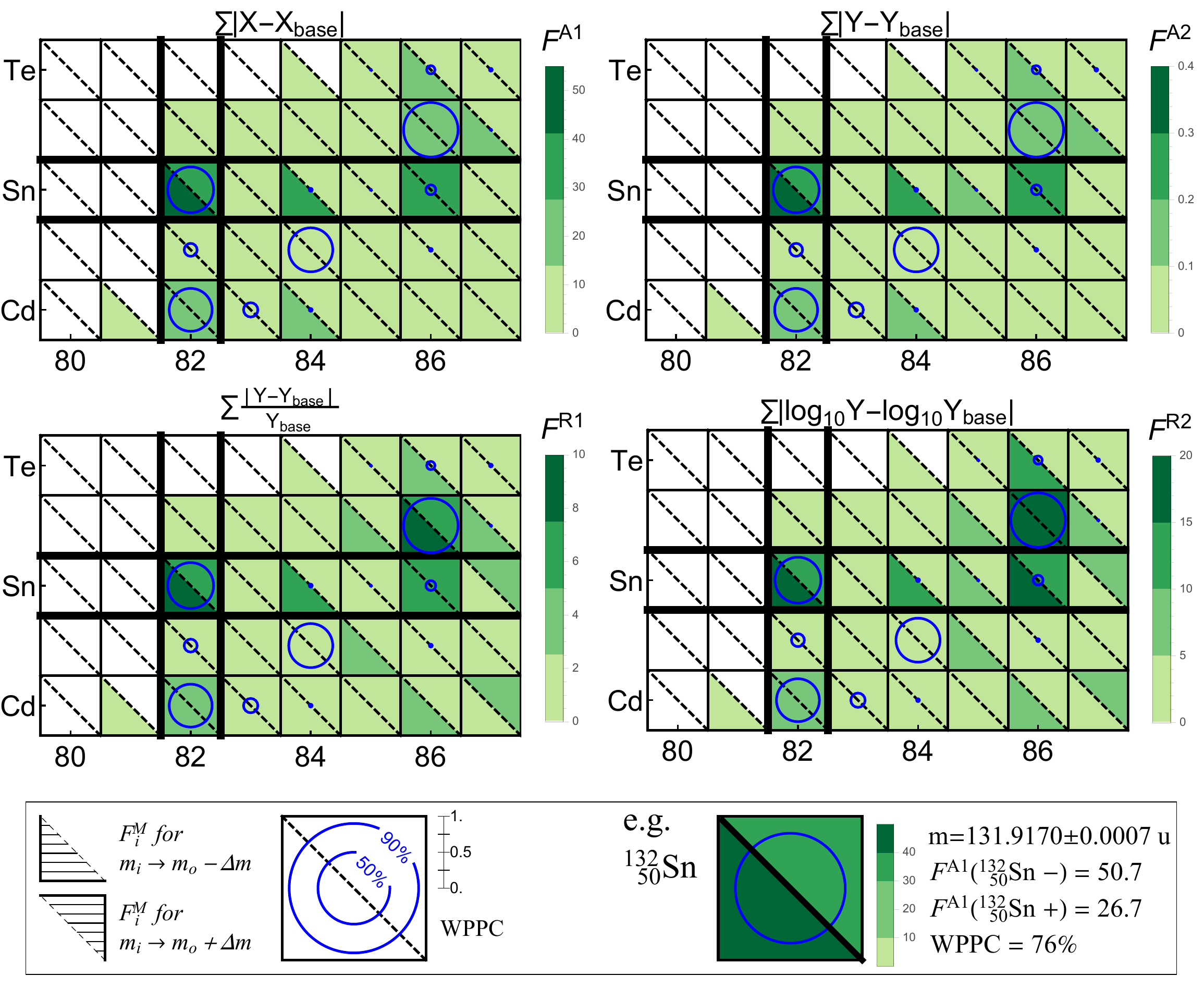}
\caption{The calculated sensitivity factors ($F^{M}_i$) for each metric is color coded as shown in
 the legend. The sensitivity factor calculated after decreasing the mass of one isotope is 
plotted in the bottom left of the isotope in the chart. Similarly the increase in mass is plotted on the top right triangle. 
The circles indicate the $r$-process waiting points population coefficients (WPPC) as determined by the nuclear Saha equation, 
where the diameter is proportional to the population coefficient where the full height of each square corresponds to 
100\%. 
The sensitivity over the mass range 
$120\le \textrm{A} \le 200$ is calculated using mass fraction normalized abundances. Even for our 
the small test study we see that the metrics disagree both qualitatively and quantitatively. \label{fig:metricCharts}}
\end{figure*}
Both global and local sensitivity metrics have been used in other works (e.g.~\citep{PhysRevC.92.035807,2012EPJA...48..184B,2012PhRvC..86c5803M}), with
global sensitivity metrics (sums of variations over mass numbers) providing a convenient and digestible value which can be used to 
estimate the total variation induced by a changing nuclear parameter (or set of parameter changes). This number alone, however, provides no 
description of how the abundances changed. For example, large overproduction of a single isotope is indistinguishable from many equivalent differences 
distributed over many isotopes. We test four different definitions of the metric (the first three have previously been used, the fourth is newly introduced in this work):
\begin{itemize}
\item A1: absolute mass fraction difference $|$X-X$_\textrm{base}|$ = $|$A (Y-Y$_\textrm{base}$)$|$
\item A2: an abundance difference $|$ Y-Y$_\textrm{base} |$
\item R1: a relative difference $|\frac{\mathrm{Y}-\mathrm{Y}_\mathrm{base}}{\mathrm{Y}_\mathrm{base}}|$
\item R2: a log-ratio $|\log_{10}\frac{\mathrm{Y}}{\mathrm{Y}_\mathrm{base}}| = | \log_{10}\mathrm{Y} - \log_{10}\mathrm{Y}_\mathrm{base}|$. 
\end{itemize}

For metrics R1 and R2 the use of mass fractions or abundances is interchangeable as the factor of A ($X = A \cdot Y$) cancels out in 
the ratio. Metrics A1 and A2 overweight changes at or near the peaks in the $r$-process distribution due to the order of magnitude differences between isotope abundances. 
They are also more sensitive to an overproduction of isotopes. In contrast, R1 and R2 do not overemphasize the more abundant isotopes which is a useful feature when examining the details of the abundance distribution. Both R1 and R2 produce similar results when the changes in ratio or percentages are small, because the series expansion to first order of the log-ratio is identical (within a constant) to the relative difference. 

\begin{figure}
\includegraphics[width=\linewidth]{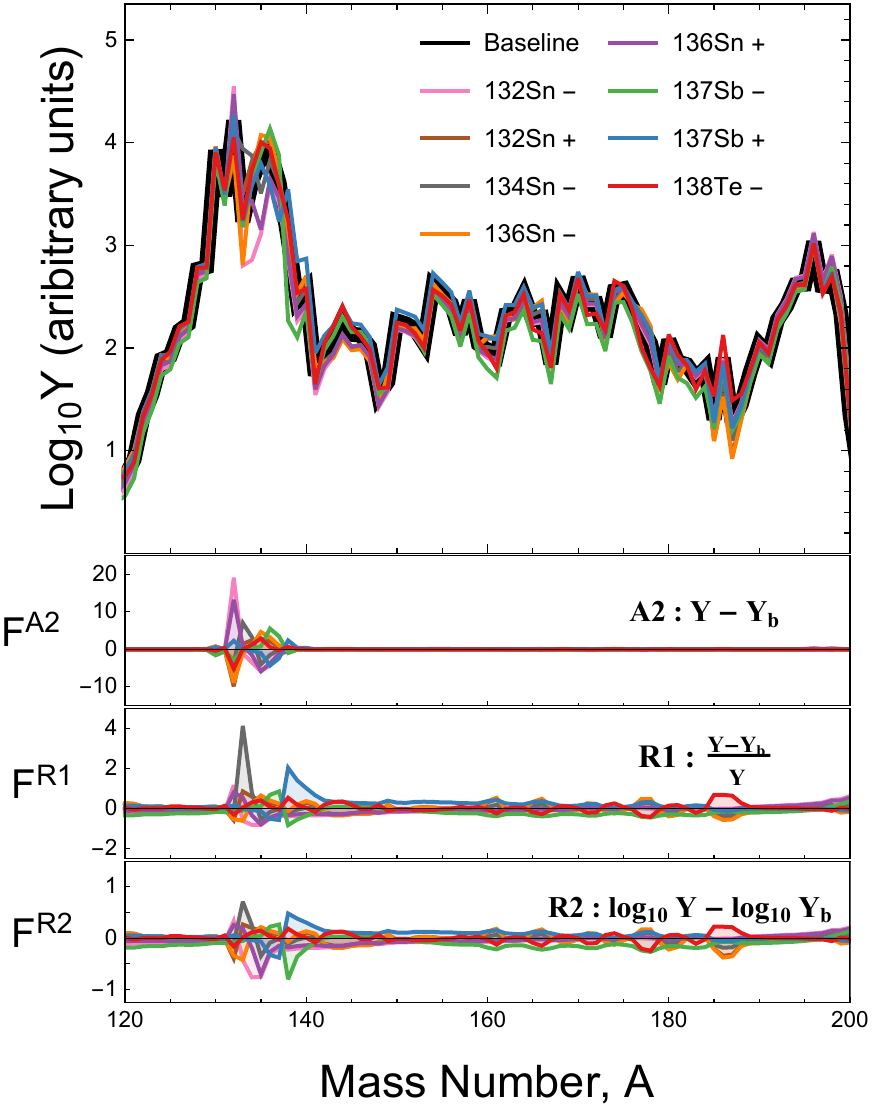}
\caption{Abundance pattern in the $r$-process simulation showing baseline vs eight most sensitive mass 
changes (labels indicate isotope's mass was changed by
$\pm0.0005\%$). The lower three panels show the deviation from the baseline as defined by the 
metrics listed in sec.~\ref{sec:metric-compare}, where the top shows metric A2 (units of mass normalized 
abundance multiplied by $10^4$), the middle shows metric R1 (unitless) and the bottom shows metric R2 (unitless). \label{fig:simulationAbundances}}
\end{figure}

Application of these metrics to the simulated data shows that each metric has varying sensitivity to the mass variations in our study. It is apparent from Fig.~\ref{fig:metricCharts}
that the four metrics have very different numerical values as expected, but significantly, they do not necessarily agree on the relative sensitivities of different isotopes. 
A1 and A2 favor the most abundant isotope $^{132}$Sn as 
expected,\footnote{Normally $^{130}$Cd is expected as the most abundance isotope in our region of interest as the precursor to the $A=130$ peak; however, in our test data set of 
the waiting point strongly favoured $^{132}$Sn at freeze-out conditions.} 
and indicate higher sensitivity to the mass of $^{132}$Sn being decreased rather than increased by the same amount.
In contrast, R1 and R2 assess $^{137}$Sb as the 
most sensitive isotope and indicate nearly identical sensitivities for the increase and decrease in 
the mass of $^{132}$Sn. Metrics A1 and A2 are identical except for a weighting by mass number present in A1 which 
will cause the two to perform differently when comparing changes near the first, second, and third abundance peaks. 
For our test data, R1 and R2 perform almost 
identically; however, R1 emphasizes overproduction of isotopes which had smaller abundances in the baseline (see $^{134}$Sn and $^{137}$Te in 
Fig.~\ref{fig:simulationAbundances} and Fig.~\ref{fig:metricCharts}) while R2 responds equally to under- and over-production by a constant factor (e.g. twice the abundance vs. half the abundance). 

In Fig.~\ref{fig:simulationAbundances} we see that the exponential sensitivity of the waiting point to the one-neutron separation 
energies ($\textrm{S}_\textrm{1n}$) leads to rather large changes to the abundances. 
While metric A2 (Y-Y$_b$) seems to imply that large changes only occur in the neighborhood of the changed nuclide, the relative metrics (R1,R2) (bottom two panels in Fig.~\ref{fig:simulationAbundances}) 
magnify changes in low-abundance nuclides, showing that there are large effects also on heavier nuclides. 
As can be seen in Fig.~\ref{fig:simulationAbundances}, changes to some isotopic masses (like $^{134}$Sn) cause 
strong local variation in the simulated abundances, while others (like $^{137}$Sb) produce 
changes to the abundance of several isobars. 
In calculating a global sensitivity factor we lose information about the details of the differences in the data, 
but greatly simplify the interpretation and presentation of the calculated sensitivity. 

Due to the differences between the metrics the best choice of metric seems unclear and depends on what type of changes in the $r$-process 
distribution are of interest. For large scale variations like the type shown in the test data here, it seems like the log-ratio R2 performs best; however, 
different changes in the nuclear parameters do not necessarily 
provoke such strong variations. Changes in the $\beta$-delayed branching ratios, for example, are likely to only cause small 
local changes during ``freeze-out,'' and in this instance A2 could be the best 
metric for studying these nuclear parameters. Due to these potential disagreements and differences in sensitivity of the 
metrics, we caution against placing too much weight on any single global sensitivity metric for assessing sensitivity to 
nuclear input parameters. 

\section{Normalization and Calibration}
\subsection{Normalization}
\label{sec:normalization-calibration}
\begin{table}
\begin{tabular}{llcccc}
  & & $F_\textrm{default}$ & $F_\textrm{min}$ &  $\Delta F $ ($\%$) & $a/b$ \\ \hline \hline
\multirow{5}{*}{$F^{(A1)}$} & $^{132}\textrm{Sn} \texttt{-}$ & 53.6 & 50.7 &  -5.4 & 0.70 \\
 & $^{136}\textrm{Sn} \texttt{+}$ & 40.2 & 39.3 &  -2.2 & 0.85 \\
 & $^{134}\textrm{Sn} \texttt{-}$ & 29.0 & 28.6 & -1.5 & 1.03 \\
 & $^{136}\textrm{Sn} \texttt{-}$ & 31.1 & 28.1 &  -9.6 & 0.81 \\
 & $^{137}\textrm{Sb} \texttt{-}$ & 26.9 & 26.9 &  -0.0 & 0.97 \\ \hline
\multirow{5}{*}{$F^{(A2)}$} & $^{132}\textrm{Sn} \texttt{-}$ & 0.396 & 0.366 &  -7.5 & 0.46 \\
 & $^{136}\textrm{Sn} \texttt{+}$ & 0.296 & 0.289 &  -2.3 & 0.82 \\
 & $^{134}\textrm{Sn} \texttt{-}$ & 0.214 & 0.210 &  -1.6 & 1.03 \\
 & $^{136}\textrm{Sn} \texttt{-}$ & 0.228 & 0.203 &  -11.0 & 0.81 \\
 & $^{132}\textrm{Sn} \texttt{+}$ & 0.223 & 0.194 &  -12.8 & 0.84 \\ \hline 
\multirow{5}{*}{$F^{(R1)}$} & $^{136}\textrm{Sn} \texttt{-}$ & 16.0 & 16.0 &  -0.0 & 1.00 \\
 & $^{132}\textrm{Sn} \texttt{+}$ & 14.9 & 14.9 &  -0.2 & 0.98 \\
 & $^{137}\textrm{Sb} \texttt{-}$ & 20.0 & 14.7 &  -26.5 & 1.24 \\
 & $^{132}\textrm{Sn} \texttt{-}$ & 15.7 & 14.7 &  -6.7 & 1.09 \\
 & $^{134}\textrm{Sn} \texttt{-}$ & 14.0 & 13.5 &  -3.5 & 1.04 \\ \hline 
\multirow{5}{*}{$F^{(R2)}$} & $^{136}\textrm{Sn} \texttt{-}$ & 7.43 & 7.27 &  -2.3 & 1.05 \\
 & $^{132}\textrm{Sn} \texttt{-}$ & 8.15 & 6.99 &  -14.2 & 1.15 \\
 & $^{132}\textrm{Sn} \texttt{+}$ & 6.73 & 6.61 &  -1.7 & 1.05 \\
 & $^{137}\textrm{Sb} \texttt{-}$ & 10.3 & 6.32 &  -38.8 & 1.31 \\
 & $^{137}\textrm{Sb} \texttt{+}$ & 6.74 & 5.57 &  -17.4 & 0.89 \\
\end{tabular}
\caption{Sensitivity factors reported using default scaling/normalization factor ($a/b = 1.0$) compared to those 
minimized sensitivity factors calculated. The blocks consist of the top five rated isotopic changes (after minimization). 
The columns list 
from left to right: the isotope whose mass was changed, the default sensitivity value, the minimal sensitivity factor, 
the percent difference, and the scaling constant which 
provided the F value. \label{table:changesFromNormalization}}
\end{table}
The $r$-process is a set of abundance values which are important in aggregate as a relative distribution and,  
without an astrophysical context, the absolute abundances are arbitrary up to a positive multiplicative constant.
When observationally available, these relative abundances are consistent with the solar $r$-process residuals 
($\textrm{N}_r = \textrm{N}_\odot - \textrm{N}_s - \textrm{N}_p$)~\cite{doi:10.1146/annurev.astro.46.060407.145207}.
In principle, a good agreement of a single simulation does not say anything: one has to run hundreds of simulations with varying astrophysical input
parameters (metallicity, stellar masses, sites, etc.) and then see the solar r-process abundances as a superposition of these values.
Observations are normalized to Ba for (s-process) or Eu (r-process), but simulated data from a single calculation is typically 
normalized by mass fraction. 
For this reason, comparing theoretical calculations to observational data or other simulation data is not so straightforward. 
In order then to assign calculate the sensitivity value
we need to somehow normalize our simulated abundances before we can calculate how sensitive the $r$-process is to
the underpinning nuclear parameter we have varied. If we had information about 
an underlying probability distribution for the $r$-process abundances it would be more obvious how to 
compare the two sets of data (e.g. $\chi^2$ or other maximum likelihood methods); however, the sparsity 
of observational data, long simulation times, and large variability 
in abundances from simulations preclude many standard statistical methods. 
For these reasons, we are forced to rely on these simple tests to estimate which inputs caused the 
largest or most significant variation.
This is possible only if we can meaningfully interpret the numerical 
values computed for our sensitivity factors and ensure that the 
methods for determining relative sensitivity are properly normalized and consistent.  

In order to understand what we mean by this, we first re-write the sensitivity metrics  
with two arbitrary scaling constants $a$ and $b$ as follows:
\begin{align}
\label{eq:F1a}
	&F_i^{(A1)} =  \sum_A \left| a X_i - b X_b \right| = \sum_A b \left| \frac{a}{b}X_i - X_b \right| \\
\label{eq:F1b}
	&F_i^{(A2)} = \sum_A \left| a Y_i - b Y_b \right| = \sum_A b \left| \frac{a}{b}Y_i - Y_b \right| \\
\label{eq:F2}
	&F_i^{(R1)} = \sum_A \left|\frac{ a Y_i - b Y_b}{b Y_b} \right| = \sum_A \left|\frac{\frac{a}{b}Y_i - Y_b}{Y_b} \right| \\
\label{eq:F3}
	&F_i^{(R2)} = \sum_A \left| \log_{10}\frac{a Y_i}{b Y_b} \right| =  |\sum_A \left| \log_{10} \frac{a}{b} Y_i - \log_{10} Y_b \right|,
\end{align}
where $Y_\textrm{base}$ is the baseline abundances and $Y_i$ is the abundance data with 
our i\textsuperscript{th} varied set of nuclear parameters.
In each metric we can identify the term $\frac{a}{b} Y_i$ which indicates that when comparing $Y_i$ to $Y_b$  we can 
find constants $a, b$ which minimize $F$. We refer to solving for the constant $a/b$ which minimizes the 
computed sensitivity factor as normalization. For our 
test data, we show the change to the final reported value for the most sensitive values (after normalization) for each metric in Table~\ref{table:changesFromNormalization}. 
The normalization procedure does not vary most values by more than a few percent for any of the metrics, but in 
the test data here the variation can be as much as 39\%. However, we do see differences which have the potential to become more pronounced 
when applied to either more comprehensive or more complex studies. In cases where there is significant non-local change, normalization 
is expected to show more significant corrections to the sensitivity factors and will allow for more robust measures of these changes.
The computational simplicity makes it 
a reasonable and prudent improvement to the existing method for computing sensitivity factors. 

\subsection{Calibration}
\begin{table}
\center
\begin{tabular}{lcccc} 
  & $F^{(A1)}$ & $F^{(A2)}$ & $F^{(R1)}$ & $F^{(R2)}$ \\ \hline \hline
 $^{130}\textrm{Cd} \texttt{-} $& 18.4 & 0.137 & 7.3  & 3.23 \\
 $^{130}\textrm{Cd} \texttt{+} $& 17.5 & 0.130 & 7.2  & 3.13 \\
 $^{131}\textrm{Cd} \texttt{-} $& 10.7 & 0.081 & 3.2  & 1.38 \\
 $^{132}\textrm{Cd} \texttt{+} $& 11.9 & 0.089 & 3.2  & 1.52 \\
 $^{135}\textrm{Cd} \texttt{+} $& 10.9 & 0.080 & 6.7  & 2.86 \\
 $^{134}\textrm{In} \texttt{-} $& 10.3 & 0.075 & 6.2  & 2.66 \\
 $^{132}\textrm{Sn} \texttt{-} $& \underline{50.7} & \underline{0.366} & 14.7 & \underline{6.99} \\
 $^{132}\textrm{Sn} \texttt{+} $& 26.7 & 0.194 & \underline{14.9} & \underline{6.61} \\
 $^{134}\textrm{Sn} \texttt{-} $& \underline{28.6} & \underline{0.210} & 13.5 & 4.73 \\
 $^{135}\textrm{Sn} \texttt{-} $& 12.4 & 0.091 & 3.9  & 1.50 \\
 $^{136}\textrm{Sn} \texttt{-} $& 28.1 & 0.203 & \underline{15.9} & \underline{7.27} \\
 $^{136}\textrm{Sn} \texttt{+} $& \underline{39.3} & \underline{0.289} & 11.4 & 5.11 \\
 $^{137}\textrm{Sn} \texttt{+} $& 11.6 & 0.085 & 7.1  & 3.04 \\
 $^{136}\textrm{Sb} \texttt{-} $& 11.0 & 0.080 & 6.4  & 2.73 \\
 $^{137}\textrm{Sb} \texttt{-} $& 26.9 & 0.193 & \underline{14.7} & 6.32 \\
 $^{137}\textrm{Sb} \texttt{+} $& 20.6 & 0.147 & 13.3 & 5.57 \\
 $^{138}\textrm{Sb} \texttt{-} $& 13.6 & 0.099 & 9.4  & 3.91 \\
 $^{138}\textrm{Te} \texttt{-} $& 17.4 & 0.127 & 10.6 & 4.51 \\ \hline
 \multicolumn{5}{c}{Range: $120 \leq A \leq 200$} \\
 \multicolumn{5}{c}{$Y_\textrm{max} = 0.165$ at $A=132$} 
\end{tabular}
\caption{Summary of global sensitivity factors computed based on different statistical metrics. The isotopes and sensitivity factors 
reported in are a combination of the top 15 most sensitive isotopes according to each metric. The 
top three most sensitive isotope mass changes are underlined in each column. These sensitivity 
values have been computed after normalization (i.e. $F^a_\textrm{min}$) according to 
each respective metric. \label{table:finalReportedValues} }
\end{table}

When writing the sensitivity factors with normalization constants, we have also explicitly exposed another potential ambiguity 
in interpreting and comparing metrics defined using simple differences of abundances (or mass fractions). 
In equations~\ref{eq:F1a} and \ref{eq:F1b} the factoring process exposed a scale parameter $b$ which pre-multiplies the 
metric. (Equations~\ref{eq:F2} and \ref{eq:F3} in contrast are purely relative or scale-free.) 
Given our arbitrary magnitude relative abundance data, differences in implicit scale ($b$) will 
cause these type of metrics to report different absolute magnitudes for sensitivity depending on 
how they happen to have been scaled. If the data is reported by an $r$-process code which relies on mass fraction normalization 
(e.g. $r$-Java 2.0) then we would expect the 2\textsuperscript{nd} $r$-process peak to have an abundance of Y$\sim10^{-3}$. 
Local changes in the peak will (for these difference type metrics) swamp out any non-local changes and the sensitivity factor 
will be close to $10^{-3}$. The values we have reported in Table~\ref{table:finalReportedValues} were computed after multiplying 
our abundances by a factor of 100 (this has the effect of making our definition of 
$F^{A1}$ in Eq.~\ref{eq:F1a} consistent with the definition in \cite{2015PhRvC..92c5807M}) and as expected, 
the reported values for $F^{(A1)}$ and $F^{(A2)}$ are the same order of magnitude as the maximum abundance
($F^{(A1)}\sim A_\textrm{max}\cdot Y_\textrm{max} = 21.8$ and $F^{(A2)}\sim Y_\textrm{max} =0.165$) as indicated 
at the bottom of Table~\ref{table:finalReportedValues}. For this reason, we recommend  
that sensitivity factors be reported with the peak abundance as we have done at the bottom of Table~\ref{table:finalReportedValues}.
The maximum abundance in the baseline, while not a perfect indicator, can 
assist in comparison of studies. 
We refer to this technique as ``calibration'' and it will allow comparison of sensitivity values from different 
studies with different baselines.

\section{Conclusion}
\label{sec:conclude}
In summary, we have performed sensitivity studies of the $r$-process using a waiting-point simulation to assess a few different global sensitivity metrics that have been proposed in recent work. While some of the biases and limitations of the metrics have been mentioned before, this is the first work that compares its various definitions by applying them on a single simulation. Besides showing that metrics of the form presented in this work give different results on the relative sensitivity of certain nuclides that lie on or near the $r$-process path, we also demonstrate that normalization of the abundances can affect the computed sensitivity. Computations done in the rare earth element region were also performed (but not presented in this manuscript) 
which confirm the preferential bias of metrics A1 and A2 for higher sensitivity near the abundance peaks. In addition, 
we presented a technique for calibrating sensitivity values from different studies. 

Based on our findings of variations from one metric to another, we recommend against their use as the only analysis tool for sensitivity. Ideally, one should understand the particular response of each metric definition,
 but computation of all metrics simultaneously (and any others that seem reasonable) seems prudent. We also recommend that the abundance distributions should be normalized in some fashion and 
 recommend reporting a minimized $F$. When reporting metrics $F^{(A1)}$ or $F^{(A2)}$, maximum abundance information should be provided to assist 
in calibration and comparison of different sensitivity studies. These recommendations are aimed at improving the strength of sensitivity studies using global sensitivity factors 
and are designed to allow for more robust statistical inference from $r$-process simulations. 

\bibliography{sensitivityPhysRevC}

\end{document}